\documentclass[12pt]{article}
\usepackage{a4wide}

\usepackage{citesort}

\usepackage{amssymb}
\usepackage{graphicx}
\usepackage{amsmath}
\usepackage[center]{subfigure}

\newcommand{\ba}{\begin{eqnarray}}
\newcommand{\ea}{\end{eqnarray}}

\newcommand{\be}{\begin{equation}}
\newcommand{\ee}{\end{equation}}

\newcommand{\bea}{\begin{eqnarray}}
\newcommand{\eea}{\end{eqnarray}}

\newcommand{\beq}{\begin{equation}}
\newcommand{\eeq}{\end{equation}}

\newcommand{\as}{a_s}

\def\ba{\begin{array}}
\def\ea{\end{array}}

\newcommand{\BreakI}{ \right. \nonumber \\ &{}& \left. }

\begin{document}
\begin{titlepage}
\begin{flushright}
TTP05-29\\
SI-HEP-2005-15\\
SFB/CPP-05-84\\
hep-ph/0512295
\end{flushright}
\vfill
\begin{center}
{\Large\bf  
Strange Quark Mass from Pseudoscalar Sum Rule 
with $O(\alpha_s^4)$ Accuracy }\\[2cm]
{\large  K.G. Chetyrkin}\,$^{a,\!}$\footnote{{\small Permanent address:
Institute for Nuclear Research, Russian Academy of Sciences,
 Moscow 117312, Russia}} 
{\large and A.~Khodjamirian}\,$^b$ \\[3mm] 
{\it $^{a}$ Institut f\"ur Theoretische Teilchenphysik, Universit\"at
Karlsruhe,\\  D-76128 Karlsruhe, Germany }\\
{\it  $^{b}$ Theoretische Physik 1, 
Fachbereich Physik, Universit\"at Siegen,\\ 
D-57068 Siegen, Germany }\\
\end{center}
\vfill

\begin{abstract}
We include the new, five-loop, $O(\alpha_s^4)$  correction
into  the QCD sum rule 
used for the $s$-quark mass determination.
The pseudoscalar Borel sum rule is taken as a study case. 
The OPE for the correlation function  with
$N^4LO$,  $O(\alpha_s^4)$  accuracy in the perturbative part, 
and with  dimension $d\leq 6$ operators reveals 
a good convergence. 
We  observe a significant improvement 
of stability  of the sum rule with respect to 
the variation of the  renormalization scale
after  including the $O(\alpha_s^4)$ correction.
We obtain the interval 
$\overline{m}_s(2 ~\mbox{GeV})=105\pm 6 \pm 7 $ MeV,
which exhibits about $ 2$ MeV increase of the central 
value, if the $O(\alpha_s^4)$ terms are removed.  

\end{abstract}
\vfill

\end{titlepage}

\section{Introduction}

A precise determination of the strange quark mass $m_s$   
is extremely important for various tests of Standard Model.
A reach variety of approaches 
is used to evaluate  this fundamental parameter 
in QCD. Recently, 
the first unquenched 
lattice QCD determinations became available  
\cite{lat_Aubin04,lat_DellaMorte05,lat_Ishikawa05,lat_Gockeler05,lat_Becirevic05,lat_Mason05}. 
In addition, ChPT provides rather accurate ratios of 
strange and nonstrange quark masses \cite{Leutw,chpt}. 
Furthermore, one evaluates $m_s$,  
combining the operator product expansion (OPE) 
of various correlation functions for strangeness-changing 
quark currents with dispersion relations. These methods include model-independent bounds \cite{bounds}, 
QCD analyses of hadronic $\tau$ decays (see 
\cite{GJPPS04,BCKtau05,Maltman:2004sk,Gorbunov:2004wy,Narison:2005zg}
for the latest results), as well as  different versions of 
QCD sum rules  \cite{SVZ} 
and related finite-energy sum rules (FESR) \cite{Chetyrkin:1978ta}.
The most recent sum rule determinations of $m_s$ 
in the channels of scalar, pseudoscalar and vector currents are presented in 
\cite{JOP02,KM01,EJS03}, respectively, references to earlier analyses can 
be found in reviews \cite{CK,Paver} (see also \cite{Narison05}).
The  estimated accuracy of these results is 15 -30\%. 
To achieve a better 
precision, one has to calculate higher orders in OPE 
of the correlation functions and gain 
a better control over potentially  important  
nonperturbative corrections beyond OPE (the so called 
''direct instantons''). Furthermore, more 
accurate data for the inputs in hadronic spectral 
functions and a better assessment of the quark-hadron 
duality are needed.
  
In this paper we concentrate on the QCD sum rules used
to evaluate the  strange quark mass  and make one further 
step to improve the accuracy of this determination  by including 
the N$^4$LO  perturbative QCD corrections of  $O(\alpha_s^4)$ 
into the sum rule. The $O(\alpha_s^4)$, five-loop
contribution has recently been calculated for the correlator
of the scalar quark currents in \cite{as4} and can be equally
well used for both scalar and  pseudoscalar sum rules.
As a study case, we choose the pseudoscalar version 
of the standard Borel sum rule. For the 
hadronic spectral function  we employ the three-resonance ansatz worked out in \cite{KM01}. 

We find that  both OPE for the correlation function
and the resulting sum rule  
reveal a good numerical convergence in powers of $\alpha_s$.
The new $O(\alpha_s^4)$ correction to the sum rule 
has a naturally small influence,
resulting in about 2 MeV decrease  of the  
s-quark mass  $\overline{m}_s$ (in $\overline{MS}$ scheme) 
determined with $O(\alpha_s^3)$ accuracy. 
Importantly, after including the $O(\alpha_s^4)$ 
correction, we observe a significant improvement in the 
stability of the extracted 
value of $\overline{m}_s$ 
with respect to the renormalization scale variation  
in the sum rule. 
With $O(\alpha_s^4)$ accuracy we obtain the interval 
\beq
\overline{m}_s(\mbox{2 GeV})= 
\left(105\pm 6\Big|_{param} \pm 7 \Big|_{hadr} \right)\mbox{MeV}\,,
\label{interval}
\eeq
where the estimated uncertainties from the 
sum rule parameters and hadronic inputs are shown separately and will be 
explained below.

In what  follows,
after a brief recapitulation of the pseudoscalar Borel sum rule
in Sect.~2, we present in Sect.~3 the QCD OPE 
expressions for the underlying correlation function,
including the  new $O(\alpha_s^4)$ terms in the 
perturbative part. The Borel transform and the 
imaginary part of the correlation function are 
also given. In Sect.~4 we turn to the numerical analysis 
of the sum rule  and obtain the interval (\ref{interval}).
Sect.~5 contains the concluding discussion. 

\section{ Pseudoscalar Sum Rule }

We consider the correlation  function: 
\beq
\Pi^{(5)}(q^2)=i\int d^4x ~e^{i q\cdot x}
\langle 0|T\left\{j_5^{(s)}(x) j^{(s)\dagger}_5(0)\right \}|0\rangle
\label{eq-corr}
\eeq
of the two  pseudoscalar strangeness-changing quark 
currents, defined as the divergences of 
the corresponding axial-vector currents: 
\beq
j_5^{(s)}=\partial^\mu (\bar{s}\gamma_\mu \gamma_5 q) =
(m_s+m_q)\bar{s}i\gamma_5 q. 
\label{eq-current}
\eeq
For definiteness, the light quark $q=u$ is taken.

The Borel sum rule is obtained following the 
standard SVZ method \cite{SVZ} and is based on the (double-subtracted) 
dispersion relation for $\Pi^{(5)}(q^2)$.
This relation is more conveniently written in a form of the 
second derivative:
\beq
\Pi^{(5)''}(q^2)\equiv \frac{d^2}{d(q^2)^2}\Pi^{(5)}(q^2)=
\frac{2}{\pi}\int\limits_0^{\infty}ds \frac{\mbox{Im}\,\Pi^{(5)}(s)}{(s-q^2)^3}\,.
\label{eq-disp}
\eeq
After Borel transformation one obtains: \footnote{Here we use 
the following normalization convention: 
${\cal B}_{M^2}[1/(a-q^2)]= e^{-a/M^2}$.} 
\beq
\Pi^{(5)''}(M^2)\equiv {\cal B}_{M^2}\left [\Pi^{(5)''}(q^2)\right ]=\frac{1}{\pi M^4}\int
\limits_0^{\infty}ds\, e^{-s/M^2}\mbox{Im}\,\Pi^{(5)}(s)\,.
\label{eq-Borel}
\eeq
The l.h.s. of the above relation is calculated in QCD 
at large $ M^ 2 \gg \Lambda_{QCD}^2$ in a form  of OPE 
(perturbative and condensate expansion), in 
powers of $\alpha_s$ and $m_s/M$, and 
up to a certain dimension of vacuum condensates. In r.h.s. the hadronic 
spectral density $\rho^{(5)}_{hadr}(s)=(1/\pi)\mbox{Im}\,\Pi^{(5)}(s)$ 
is substituted. At $s<s_0$ , where $s_0$ is some effective threshold,
the function $\rho^{(5)}_{hadr}(s)$ includes kaon
and its excitations.   
The rest of the hadronic dispersion integral 
at $s>s_0$  is approximated using quark-hadron duality,
$\rho^{(5)}_{hadr}(s)\simeq \rho^{(5)}_{OPE}(s)$, 
with the spectral function calculated from OPE: 
$ \rho^{(5)}_{OPE}(s)=(1/\pi) \mbox{Im}\,[\Pi^{(5)}(s)]_{OPE}$.  
The final form of the sum rule is:
\beq
M^4[\Pi^{(5)''}(M^2)]_{OPE}=
\int \limits_0^{s_0}ds\,  e^{-s/M^2}\rho^{(5)}_{hadr}(s) +
\int\limits_{s_0}^{\infty}ds\, e^{-s/M^2}\rho^{(5)}_{OPE}(s)\,.
\label{SR}
\eeq
In the following we discuss both parts of this equation
in detail.

\section{OPE results  to $O(\alpha_s^4)$ }

In this section we present the expressions for $[\Pi^{(5)''}(q^2)]_{OPE}$ 
and, correspondingly, for $[\Pi^{(5)''}(M^2)]_{OPE}$ and $ \rho^{(5)}_{OPE}(s)$ 
determining the QCD input in the sum rule (\ref{SR}). 
The OPE for $[\Pi^{(5)''}(q^2)]_{OPE}$ goes over powers of 
$(1/q^2)^{d+2}$ ordered  by the dimension $d=0,2,4,6$. The OPE terms
with $d>6$ are neglected, while already the $d=6$ contribution 
is very small in the working region of the variables $Q^2$ and  $M^2$.   

The $d=0,2$ terms of OPE originate from the 
perturbative part of the correlation function. The expansion
in quark-gluon coupling  up to four loops, that is, up to  $O(\alpha_s^3)$, 
can be taken from \cite{Gorishnii,Chetyrkin:1996sr,CPS96}. The new $O(\alpha_s^4)$
terms are obtained in \cite{as4}. Putting them together, we obtain:
\beq
[\Pi^{(5)''}(Q^2)]_{OPE}^{(d=0,2)}
=\frac{3(m_s+m_u)^2}{8\pi^2Q^2}\Bigg\{1+
\sum_i\bar{d}_{0,i}\, \as^i 
-2\frac{m_s^2}{Q^2}
\left(1+\sum_i\bar{d}_{2,i} \, \as^i 
\right)
\Bigg\}
\label{piOPE} 
{},
\eeq
where $Q^2=-q^2$, and the coefficients multiplying the 
powers of the quark-gluon coupling 
$a_s=\alpha_s(\mu)/\pi$ are
\begin{eqnarray}
{\bar{d}_{0,1} =   
\frac{11}{3} 
-2  \,l_{Q}
%zero == 0
{},
}
\  \
\bar{d}_{0,2} =   
\frac{5071}{144} 
-\frac{35}{2}  \,\zeta_{3}
-\frac{139}{6}  \,l_{Q}
+\frac{17}{4}  \,l_Q^2
{},
\label{d01_d02}
\end{eqnarray}
%zero == 0
\begin{eqnarray}
\bar{d}_{0,3} =   
\frac{1995097}{5184} 
-\frac{1}{36}  \pi^4
-\frac{65869}{216}  \,\zeta_{3}
+\frac{715}{12}  \,\zeta_{5}
-\frac{2720}{9}  \,l_{Q}
+\frac{475}{4}  \,\zeta_{3} \,l_{Q}
+\frac{695}{8}  \,l_Q^2
-\frac{221}{24}  \,l_Q^3
%zero == 0
{},
\label{d03}
\end{eqnarray}
%zero == 0
\begin{eqnarray}
{\bar{d}_{0,4} =  } 
&{}&
\frac{2361295759}{497664} 
-\frac{2915}{10368}  \pi^4
-\frac{25214831}{5184}  \,\zeta_{3}
+\frac{192155}{216}  \,\zeta_3^2
+\frac{59875}{108}  \,\zeta_{5}
-\frac{625}{48}  \,\zeta_{6}
%zero == 0
\nonumber\\
&{-}& \frac{52255}{256}  \,\zeta_{7}
+l_{Q}\left[
-\frac{43647875}{10368} 
+\frac{1}{18}  \pi^4
+\frac{864685}{288}  \,\zeta_{3}
-\frac{24025}{48}  \,\zeta_{5}
%zero == 0
\right]
\nonumber\\
&{+}& \,l_Q^2
\left[
\frac{1778273}{1152} 
-\frac{16785}{32}  \,\zeta_{3}
%zero == 0
\right]
{+} \,l_Q^3
\left[
-\frac{79333}{288}\right]
{+} \,l_Q^4
\left[
 \frac{7735}{384}\right]
{},
\label{d04}
\end{eqnarray}
%zero == 0

\begin{eqnarray}
\bar{d}_{2,1} =
 \frac{28}{3}
-4 \,l_{Q}
{},
\ \
\bar{d}_{2,2} =
\frac{8557}{72}
-\frac{77}{3}  \,\zeta_{3}
%zero == 0
-\frac{147}{2}
\,l_{Q}
+ \,\frac{25}{2}
 \,l_Q^2
{}\,,
\label{d21+d22}
\end{eqnarray}
%zero == 0
%zero == 0
including the new result for $\bar{d}_{0,4}$.
Here $l_Q= \log\frac{Q^2}{\mu^2}$, and $\zeta_n\equiv \zeta(n)$ is the Riemann's Zeta-function.  The coupling $a_s$  
and the quark masses $m_s$ and $m_u$ are 
all taken  in $\overline{MS}$ scheme at the renormalization scale $\mu$. 
We have neglected the light-quark mass $m_u$, 
except in the overall factors. Note also that in the 
subleading $d=2$, $O(m_s^4)$ terms of the above expansion, the currently 
achieved  $O(\alpha_s^2)$ accuracy is quite sufficient. 

The contributions with $d=4,6$ in the correlation function 
originate both from nonperturbative (condensate) terms  and from $O(m_s^6)$
corrections, and we use the known expressions \cite{JM,CPS96}  
\bea
&{}&[\Pi^{(5)''}(q^2)]_{OPE}^{(d=4,6)}=
 \frac{(m_s+m_u)^2}{Q^6}\left\{-2m_s\langle \bar uu\rangle
\left(1 + a_s(\frac{23}{3}-2l_Q)\right)
\right.
\nonumber \\&{} &\left. 
-\frac19 I_G\left(1 + a_s(\frac{121}{18} - 2l_Q)\right)
%\\&{} &\left. 
+ I_s\left(1+a_s(\frac{64}{9}-2l_Q)\right)
\nonumber \right. \\&{} &\left. 
-\frac{3}{7\pi^2}m_s^4\left(\frac{1}{a_s}+
\frac{155}{24} - \frac{15}{4}l_Q \right)
+\frac{I_6}{Q^2} \right\}\,,
\label{piOPEnp}
\eea
where 
\be
I_s=m_s\langle \bar{s}s\rangle+\frac{3}{7\pi^2}
m_s^4\left(\frac{1}{a_s}-\frac{53}{24}\right )
\ee
and 
\be
I_G=-\frac{9}{4}\langle \frac{\alpha_s}{\pi}
G^2\rangle\left (1+\frac{16}{9}a_s\right)+
4a_s\left(1+\frac{91}{24}a_s\right)m_s\langle \bar{s}s\rangle+\frac{3}{4\pi^2}\left(1+\frac{4}{3}a_s\right)
m_s^4
\ee
are the vacuum expectation values of two RG-invariant combinations  of
dimension 4 containing quark and gluon condensate densities 
(for details and explanation see
\cite{Spiridonov:1988md,Chetyrkin:1994qu,JM,CPS96}). Finally, 
\be
I_6= -3m_s\langle\bar{u}uG\rangle-
\frac{32}9\pi^2a_s\Big (\langle\bar{u}u\rangle^2
+\langle\bar{s}s\rangle^2-9\langle\bar{u}u\rangle
\langle\bar{s}s\rangle\Big)
\ee
is the combination of dimension-6 contributions 
of the quark-gluon and 4-quark condensates
(the vacuum saturation is assumed for the latter).

The Borel transform of Eqs.~(\ref{piOPE}) and (\ref{piOPEnp}) 
is given by 
\beq
[\Pi^{(5)''}(M^2)]_{OPE}^{(d=0,2)}=
\frac{3(m_s+m_u)^2}{8\pi^2}\Bigg\{1+
\sum_i\bar{b}_{0,i}\, \as^i 
- 
2 
\frac{m_s^2}{M^2}
\left(
1
+
\sum_i\bar{b}_{2,i} \, \as^i 
\right)
\Bigg \} 
\label{BpiOPE} 
{},
\eeq
where $l_M = \log\frac{M^2}{\mu^2}$ and the coefficients are
\begin{eqnarray}
{\bar{b}_{0,1} =  } 
\frac{11}{3} 
+2  \, \gamma_E
-2  \,l_{M}
%zero == 0
{},
\label{b01}
\end{eqnarray}
%zero == 0
\begin{eqnarray}
{\bar{b}_{0,2} =  } 
\frac{5071}{144} 
+\frac{139}{6}  \, \gamma_E
+\frac{17}{4}  \,\gamma_E^2
-\frac{17}{24}  \pi^2
-\frac{35}{2}  \,\zeta_{3}
-\frac{139}{6}  \,l_{M}
-\frac{17}{2}  \, \gamma_E \,l_{M}
+\frac{17}{4}  \,l_M^2
%zero == 0
{},
\label{b02}
\end{eqnarray}
%zero == 0
\begin{eqnarray}
\bar{b}_{0,3} &=&   
\frac{1995097}{5184} 
+\frac{2720}{9}  \, \gamma_E
+\frac{695}{8}  \,\gamma_E^2
+\frac{221}{24}  \,\gamma_E^3
-\frac{695}{48}  \pi^2
-\frac{221}{48}  \, \gamma_E \pi^2
\nonumber
\\
&{}&
\phantom{+}
-\frac{1}{36}  \pi^4
-\frac{61891}{216}  \,\zeta_{3}
-\frac{475}{4}  \, \gamma_E \,\zeta_{3}
+\frac{715}{12}  \,\zeta_{5}
%zero == 0
\nonumber\\
&{+}& \,l_{M}
\left[
-\frac{2720}{9} 
-\frac{695}{4}  \, \gamma_E
-\frac{221}{8}  \,\gamma_E^2
+\frac{221}{48}  \pi^2
+\frac{475}{4}  \,\zeta_{3}
%zero == 0
\right]
\nonumber\\
&{+}& \,l_M^2
\left[
\frac{695}{8} 
+\frac{221}{8}  \, \gamma_E
%zero == 0
\right]
-\frac{221}{24} \,l_M^3
{},
\label{b03}
\end{eqnarray}
%zero == 0
%%%%%%%%%%%%%%%%%%%%%%%%%%%%%%%%%%
\begin{eqnarray}
b_{0,4} &=&   
\left.
\frac{2361295759}{497664} 
+\frac{43647875}{10368}  \, \gamma_E
+\frac{1778273}{1152}  \,\gamma_E^2
+\frac{79333}{288}  \,\gamma_E^3
+\frac{7735}{384}  \,\gamma_E^4
-\frac{1778273}{6912}  \pi^2
\BreakI
\phantom{+}
-\frac{79333}{576}  \, \gamma_E \pi^2
-\frac{7735}{384}  \,\gamma_E^2 \pi^2
+\frac{2263}{41472}  \pi^4
-\frac{1}{18}  \, \gamma_E \pi^4
-\frac{22358843}{5184}  \,\zeta_{3}
-\frac{818275}{288}  \, \gamma_E \,\zeta_{3}
\BreakI
\phantom{+}
-\frac{16785}{32}  \,\gamma_E^2 \,\zeta_{3}
+\frac{5595}{64}  \pi^2 \,\zeta_{3}
+\frac{192155}{216}  \,\zeta_3^2
+\frac{59875}{108}  \,\zeta_{5}
+\frac{24025}{48}  \, \gamma_E \,\zeta_{5}
-\frac{625}{48}  \,\zeta_{6}
-\frac{52255}{256}  \,\zeta_{7}
%zero == 0
\right.
\nonumber\\
&{+}& \,l_{M}
\left[
-\frac{43647875}{10368} 
-\frac{1778273}{576}  \, \gamma_E
-\frac{79333}{96}  \,\gamma_E^2
-\frac{7735}{96}  \,\gamma_E^3
+\frac{79333}{576}  \pi^2
+\frac{7735}{192}  \, \gamma_E \pi^2
\BreakI
\phantom{+ \,l_{M}}
+\frac{1}{18}  \pi^4
+\frac{818275}{288}  \,\zeta_{3}
+\frac{16785}{16}  \, \gamma_E \,\zeta_{3}
-\frac{24025}{48}  \,\zeta_{5}
%zero == 0
\right]
\nonumber\\
&{+}& \,l_M^2
\left[
\frac{1778273}{1152} 
+\frac{79333}{96}  \, \gamma_E
+\frac{7735}{64}  \,\gamma_E^2
-\frac{7735}{384}  \pi^2
-\frac{16785}{32}  \,\zeta_{3}
%zero == 0
\right]
\nonumber\\
&{+}& \,l_M^3
\left[
-\frac{79333}{288} 
-\frac{7735}{96}  \, \gamma_E
%zero == 0
\right]
+ \,l_M^4
\left[
 \frac{7735}{384}
\right]
{},
\label{b04}
\end{eqnarray}
%zero == 0

%%%%%%%%%%%%%%%%%%%%%%%%%%%%%%%%%%%%
\begin{eqnarray}
{\bar{b}_{2,1} =  } 
\frac{16}{3} 
+4  \, \gamma_E
-4  \,l_{M}
%zero == 0
{},
\label{b21}
\end{eqnarray}
%zero == 0
\begin{eqnarray}
\bar{b}_{2,2} =  
\frac{5065}{72} 
+\frac{97}{2}  \, \gamma_E
+\frac{25}{2}  \,\gamma_E^2
-\frac{25}{12}  \pi^2
-\frac{77}{3}  \,\zeta_{3}
-\frac{97}{2}  \,l_{M}
-25  \, \gamma_E \,l_{M}
+\frac{25}{2}  \,l_M^2
%zero == 0
{},
\label{b22}
\end{eqnarray}
%zero == 0and
and, respectively, 
\bea
[\Pi^{(5)''}(M^2)]_{OPE}^{(d=4,6)} &=&
\frac{(m_s+m_u)^2}{2M^4}\left\{-2m_s\langle \bar uu\rangle
\left(1 + a_s(\frac{14}{3}+2\gamma_E-2l_M)
\right)
\right.
\nonumber
\\
& &\left.
\hspace{-2cm} -\frac19 I_G\left(1 + a_s(
\frac{67}{18} + 2\gamma_E-2l_M)\right)
+ I_s\left(1+ a_s(\frac{37}{9}+2\gamma_E-
2l_M)\right) 
\right.\nonumber\\
& &\left.
\hspace{-2cm} -\frac{3}{7\pi^2}m_s^4\left(\frac{1}{a_s}+
\frac{5}{6} + \frac{15}{4}\gamma_E-\frac{15}{4}l_M
\right)+\frac{I_6}{3M^2}\right\}
\,.\label{BpiOPEnp} 
\eea
In addition, we need the imaginary part of the 
correlation function calculated 
with the same $\alpha_s^4$ accuracy as Eqs.~(\ref{BpiOPE}) 
and (\ref{BpiOPEnp}):
\bea
\rho^{(5)}_{OPE}(s) &=&\frac{1}{\pi}\mbox{Im}\Pi^{(5)}(s)=
\frac{3(m_s+m_u)^2}{8\pi^2}\,s\,\Bigg\{1+
\sum_i\tilde{r}_{0,i}\, \as^i 
- 
2 \frac{m_s^2}{s}
\left(
1+\sum_i\tilde{r}_{2,i} \, \as^i 
\right)
\Bigg \} 
\nonumber\\
%\hspace*{-1.5cm}
&+&
 \frac{m_s^2(s)}{s}\left\{ \frac{45}{56\pi^2}m_s^4(s)
 +
2a_s(s)m_s\langle \bar uu\rangle +
\frac{a_s(s)}{9}I_G - a_s(s)I_s\right\}
%\nonumber
\,,
\label{ImpiOPE}
\eea
where $l_s = \log\frac{s}{\mu^2}$ and 
\begin{eqnarray}
\tilde{r}_{0,1} =  
\frac{17}{3} 
-2  \, l_s,
\ \ 
{\tilde{r}_{0,2} =  } 
\frac{9631}{144} 
-\frac{17}{12}  \pi^2
-\frac{35}{2}  \,\zeta_{3}
-\frac{95}{3}  \, l_s
+\frac{17}{4}  \,l_s^2
%zero == 0
{},
\label{r02}
\end{eqnarray}

%zero == 0
\begin{eqnarray}
\tilde{r}_{0,3} =  
&{}&
\frac{4748953}{5184} 
-\frac{229}{6}  \pi^2
-\frac{1}{36}  \pi^4
-\frac{91519}{216}  \,\zeta_{3}
+\frac{715}{12}  \,\zeta_{5}
-\frac{4781}{9}  \, l_s
+\frac{221}{24}  \pi^2 \, l_s
\nonumber
\\
&{}&
\phantom{+}
+\frac{475}{4}  \,\zeta_{3} \, l_s
+\frac{229}{2}  \,l_s^2
-\frac{221}{24}  \,l_s^3
%zero == 0
{},
\label{r03}
\end{eqnarray}
%zero == 0
\begin{eqnarray}
\tilde{r}_{0,4} =  
&{}&
\frac{7055935615}{497664} 
-\frac{3008729}{3456}  \pi^2
+\frac{19139}{5184}  \pi^4
-\frac{46217501}{5184}  \,\zeta_{3}
+\frac{5595}{32}  \pi^2 \,\zeta_{3}
\nonumber
\\
&{}&
\phantom{+}
+\frac{192155}{216}  \,\zeta_3^2
+\frac{455725}{432}  \,\zeta_{5}
-\frac{625}{48}  \,\zeta_{6}
-\frac{52255}{256}  \,\zeta_{7}
%zero == 0
\nonumber\\
&{+}& \, l_s
\left[
-\frac{97804997}{10368} 
+\frac{51269}{144}  \pi^2
+\frac{1}{18}  \pi^4
+\frac{1166815}{288}  \,\zeta_{3}
-\frac{24025}{48}  \,\zeta_{5}
%zero == 0
\right]
\nonumber\\
&{+}& \,l_s^2
\left[
\frac{3008729}{1152} 
-\frac{7735}{192}  \pi^2
-\frac{16785}{32}  \,\zeta_{3}
%zero == 0
\right]
%+ \,l_s^3
-\frac{51269}{144}\,l_s^3
{+}
% \,l_s^4
 \frac{7735}{384}\,l_s^4
{},
\label{r04}
\end{eqnarray}
%zero == 0
\begin{eqnarray}
\tilde{r}_{2,1} =  
\frac{16}{3} 
-4  \, l_s
{},
\ \ 
\tilde{r}_{2,2} =  
\frac{5065}{72} 
-\frac{25}{6}  \pi^2
-\frac{77}{3}  \,\zeta_{3}
-\frac{97}{2}  \, l_s
+\frac{25}{2}  \,l_s^2
%zero == 0
{}.
\label{r21_22}
\end{eqnarray}
%zero == 0

The OPE expressions are valid at sufficiently large 
$Q^2\gg \Lambda_{QCD}^2$ or, correspondingly,  at large
$M^2$. It is well known that in the spin zero (scalar and pseudoscalar) 
channels the breakdown of OPE is expected to occur at 
relatively large 
$Q^2 \simeq $ 1 GeV$^2$, due to the presence of nonperturbative
vacuum effects which are beyond the local
condensate expansion \cite{inst,NSVZ}. 
Models of the correlation function based on instanton ensembles,
such as the instanton liquid model (ILM) \cite{ILM,ILM1} 
allow to penetrate to smaller $Q^2$. A remedy used in previous analyses of 
pseudoscalar sum rules is to 
add to the OPE series an instanton correction calculated in ILM.
As realized, e.g., in \cite{KM01}, at sufficiently large $M^2$, 
practically already at $M^2>$ 2 GeV$^2$ the ILM correction is small, 
hence  we will avoid it  by choosing  2 GeV$^2$ as a lower limit of 
the Borel mass.

For the reader's  convenience, the  lengthy 
coefficients appearing
in eqs.~(\ref{piOPE}), (\ref{BpiOPE}) and (\ref{ImpiOPE})  
are made  available (in  computer-readable form) in 
\cite{comp}.

\section{Hadronic spectral density and the sum rule}

The spectral function $\rho^{(5)}_{hadr}(s)$ 
in Eq.~(\ref{SR}) is a positive definite sum of all hadronic states 
with strangeness and $J^P=0^-$, located
below the threshold $\sqrt{s_0}$, above which 
$\rho^{(5)}_{hadr}(s)$ is approximated by the OPE spectral density.
Clearly, the larger is $s_0$, 
the smaller is the sensitivity of the sum rule
to this quark-hadron duality ansatz.  

The lowest hadronic state is the kaon.
Using the standard definition of the kaon decay constant 
\beq
\langle 0 |\bar{s}\gamma_\mu\gamma_5 u|K^+(q)\rangle
= i q_\mu f_K\,,
\label{eq-fKax}
\eeq
one obtains  the relevant hadronic matrix element of the pseudoscalar 
current:
\beq
\langle 0 |j_5^{(s)}|K^+(q)\rangle = f_K m_K^2\,,
\label{eq-fKdef}
\eeq
so that the kaon contribution to the hadronic 
spectral density reads:
\beq
\rho^{(5)}_{K}(s)= f_K^2m_K^4\delta(m_K^2-s)\,.
\eeq
The two heavier pseudoscalar resonances \cite{PDG} 
are  $K_1 = K(1460)$ and  $K_2=K(1830)$ with the masses 
$m_{K_1}= 1460$ MeV and $m_{K_2}= 1830$ MeV and 
total widths $\Gamma_{K_1}= 260$ MeV
and $\Gamma_{K_2}= 250 $ MeV, respectively. 
These resonances are not yet well established, in particular, 
no experimental errors are attributed to their masses  and widths.  
In any case, it seems plausible 
that the hadronic spectral density in the pseudoscalar 
channel with strangeness is dominated by the kaon 
and $K_{1,2}$ resonances, 
making this channel less complicated than 
the scalar channel where the strong $K\pi$
scattering in S-wave ($J^P=0^+$)
demands a dedicated analysis (see e.g., \cite{CDNP97,JOP02}).

A  detailed analysis of the hadronic part in the 
pseudoscalar sum rules (in both FESR and Borel 
versions) is presented in \cite{KM01},
employing the hadronic spectral density where the 
contributions of two resonances 
$K_{1,2}$ with finite widths are simply added to the ground-state 
term of the kaon. 
Here we adopt the same ansatz for the 
hadronic spectral density 
\footnote{For a different hadronic ansatz 
including  $K^* \pi$ state explicitly, see \cite{DPS97}.} in the sum rule 
(\ref{SR}):
\beq
\rho^{(5)}_{hadr}(s)=f_K^2m_K^4\delta(m_K^2-s)+\sum\limits_{i=1,2}f_{K_i}^2 m_{K_i}^4 B_{K_i}(s)\,,
\label{eq-hadr_spectr}
\eeq
where $B_{K_i}(s)$ are the 
finite-width (Breit-Wigner type)
replacements of  the $\delta$-function
in the spectral density for $K_{1,2}$:
\beq
\delta(m_{K_{i}}^2-s)\to B_{K_i}(s)=
\frac{1}{\pi}\left (\frac{\Gamma_{K_i}m_{K_i}}
{(s-m_K^2)^2+(\Gamma_{K_i}m_{K_i})^2}\right)\, . 
\label{BW}
\eeq
In \cite{KM01} using FESR,  
the decay constants $f_{K_1}$ and $f_{K_2}$ of $K_1$ and $K_2$ resonances
(defined similarly to $f_K$) were fitted. As anticipated from 
ChPT, small values, in the ballpark of 20-30 MeV    
for both $f_{K_1}$ and $f_{K_2}$   
were obtained. On the other hand, due to 
the large mass multiplying these constants, the effects 
of $K_{1}$ and $K_2$ are quite noticeable in the hadronic part of the sum rule,
hence, one has to avoid too large values of $M^2$.  
We will use the estimates of $f_{K_1,K_2}$ from \cite{KM01} 
as hadronic inputs in our numerical analysis of (\ref{SR}).

Further improvements of the hadronic ansatz are possible, but they 
our beyond our scope here. 
In particular, it seems important
to investigate the role of multiparticle states 
in the hadronic spectral function, 
starting from the two-particle states $K^* \pi, K\rho$.
In \cite{KM01} it is assumed that multiparticle effects are 
at least partially 
taken into account in the finite widths of $K_{1,2}$. 
One usually neglects the possible contributions 
of the nondiagonal transitions to the hadronic 
spectral function, e.g., intermediate states  
of the type $\langle 0| j_5^{(s)} |K \rangle \langle K| K^*\pi\rangle 
\langle K^*\pi|K_1\rangle \langle K_1|j_5^{(s)}|0\rangle $. 
The analysis of the light-quark 
vector channel ($J^{P}=1^-$) without and with strangeness 
(see e.g.,\cite{BKK}) indicates that the effects of mixing  between 
separate resonances via intermediate multiparticle states
could be noticeable.  Here, adopting the ansatz 
 (\ref{eq-hadr_spectr}) we tacitly assume that the 
total widths of $K_{1,2}$ account for the dominant contributions
of multiparticle states. In order to estimate the influence
of this effect, we will also consider a version of the hadronic spectral density 
(\ref{eq-hadr_spectr}) with the total widths of $K_{1,2}$ set to zero, 
interpreting the difference 
of the result with and without the widths as a rough estimate 
of the uncertainty due to multiparticle hadronic states.

\section{Inputs and  numerical results}

For the running of the strong coupling $a_s$ and of 
the quark masses in $\overline{MS}$ 
scheme we use the four-loop approximation  and employ 
the numerical program RunDec described in \cite{RunDec}.
The reference value for the quark-gluon coupling 
is taken as $\alpha_s(m_Z)= 0.1187$ \cite{PDG}.
The alternative choice $\alpha_s(m_\tau)= 0.334$ 
\cite{PDG} produces a small difference
which we include into the overall counting 
of uncertainties.    
We do not attempt to fit the $u$- and $d$-quark masses
from the analogous sum rules, and simply take 
the current (non-lattice) intervals from \cite{PDG}:
$\overline{m}_u (2~\mbox{GeV})=(1.5 ~\mbox{-}~ 5.0) ~\mbox{MeV}$, $\overline{m}_d (2~\mbox{GeV})=
(5.0 ~\mbox{-}~ 9.0) ~\mbox{MeV}$. 

The renormalization scale in our numerical calculation is taken as $\mu=M$, 
reflecting the average virtuality
of perturbative quarks and gluons in the correlator.
In order to study the scale dependence we also vary the scale 
within $M^2/2 <\mu^2 < 2M^2$. The window of Borel parameter
is taken as in \cite{KM01}, $2 <M^2<3 $ GeV$^2$. This choice
allows to avoid large nonperturbative effects, 
simultaneously keeping  the excited state contributions 
reasonably small.  

The remaining input parameters used for the OPE of the 
correlation function are: 
the quark condensate densities
taken from GMOR relation 
$\langle \bar{u} u\rangle=
-f_\pi^2 m_\pi^2/(2(m_u+m_d))$
where $f_\pi=130.7 $ MeV \cite{PDG}; the ratio of strange and 
nonstrange condensates ~~ 
$ \langle \bar{s} s\rangle/\langle \bar{u} u\rangle= 0.8 \pm 0.3$; 
the gluon condensate density  
$\langle \alpha_s/\pi GG\rangle= (0.012^{+ 0.006}_{-0.012})~\mbox{GeV}^4$.
Finally, the dimensionful parameter for the quark-gluon 
condensate density is taken as $m_0^2=0.8\pm 0.2$ GeV$^2$  
(for a recent comprehensive review of condensates see \cite{Ioffe05}).

First of all, we  address the main question which interests 
us here, namely, how good is the convergence of
OPE for the correlation function in $N^4LO$, and how large 
is the numerical impact of the new $O(\alpha_s^4)$ correction. 
For that we define the ratios:
\begin{equation}
r_n^{(d=0,2)}(M^2)=
\frac{\{[\Pi^{(5)''}(M^2)]_{OPE}^{(d=0,2)}\}^{O(\alpha_s^n)}}{[\Pi^{(5)''}(M^2)]_{OPE}^{(d=0,2)}+
[\Pi^{(5)''}(M^2)]_{OPE}^{(d=4,6)}}
\label{eq-rn}
\end{equation}
for $n=0,1,2,3,4$, 
where the numerator contains the contribution 
of $O(\alpha_s^n)$ to the Borel transformed 
correlation function. 
The analogous ratio for the nonperturbative contributions is 
$r^{(d=4,6)}(M^2)$, where the numerator 
contains only the power suppressed term 
$[\Pi^{(5)''}(M^2)]_{OPE}^{(d=4,6)}$. Altogether,

$$\sum_{n=0,1,2,3,4} r_n^{(d=0,2)}+r^{(d=4,6)}=1.$$
Note that the dominant $m_s$-dependence in $\Pi^{(5)''}(M^2)$ is 
due to the overall factor $(m_s+m_u)^2$ and 
largely cancels in $r_n^{(d=0,2)}$ and $r^{(d=4,6)}$. (We use
$\overline{m}_s(2 \mbox{GeV})=105$ MeV
in the suppressed terms for this numerical illustration).
In Fig.~1 the ratios $ r_n^{(d=0,2)}$ and $r^{(d=4,6)}$ are plotted
as a function of Borel parameter squared. The convergence is excellent,
even beyond the region of the Borel parameter 
chosen for the sum rule analysis. In  
the central point  $M^2=2.5$ GeV$^2$  
we obtain $r_n^{(d=0,2)}(2.5 ~\mbox{GeV}^2) =
{52.4 \%, 28.3 \%, 14.4\%, 4.0\% ,-0.3 \%}$
for  $n=0,1,2,3,4$, respectively and 
$r^{(d=4,6)}(2.5 ~\mbox{GeV}^2)=1.2 \% $. We conclude that 
the currently achieved  accuracy of the correlation
function at large $M^2$ is quite sufficient for 
the applications, such as  the quark mass determination.

%%%%%%%%%%%%%%%%%%%%%FIG.1 %%%%%%%%%%%%%%%%%%
\begin{figure}[t]
\begin{center}
{\includegraphics[scale=0.7]{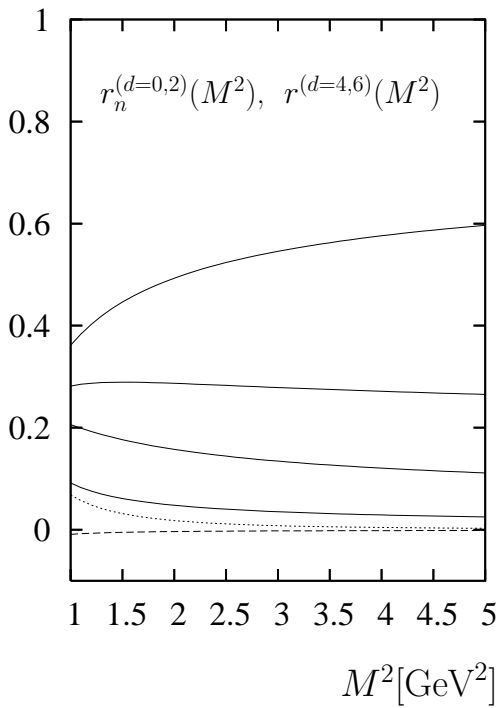}}
\end{center}
\caption{\it Relative contributions to OPE 
of the correlation function $\Pi^{(5)''}(M^2)$
defined in (\ref{eq-rn}),
plotted as functions of the Borel parameter squared.
The solid lines from up to down correspond to $r_n^{(d=0,2)}$ with $n=0,1,2,3$,
respectively, the dashed line to $r_4^{(d=0,2)}$ and  the dotted line 
to $r^{(d=4,6)}$.
}
\label{fig1}
\end{figure}

We then turn to the sum rule (\ref{SR}).  
The input parameters for the 
kaon contribution to the hadronic part are: $f_K=159.8 ~\mbox{MeV}$, 
$m_K=493.7$ MeV \cite{PDG}.
The masses and total widths of $K_1$ and $K_2$ 
resonances \cite{PDG} were already quoted in the previous section. 
Their decay constants are taken from \cite{KM01}: 
$f_{K_1}/\sqrt{2}=(22.9 \pm 2.4) ~\mbox{MeV}$,
$f_{K_2}/\sqrt{2}=(14.5 \pm 1.5) ~\mbox{MeV}$,
where the factor $1/\sqrt{2}$ accounts for the 
difference between the normalizations. Note that 
for consistency, we take the version of \cite{KM01} 
obtained without ILM correction; furthermore, we
added the uncertainties of $f_{K_1},f_{K_2}$ given in \cite{KM01} 
in quadrature. The duality threshold  adopted in our 
calculation is $s_0=4.5\pm 0.5~\mbox{GeV}^2$. 
The central value provides the best stability 
of the sum rule in the Borel parameter interval 2-3 GeV$^ 2$,
whereas the spread of $\pm 0.5$ GeV$^2$
is added to  allow for some additional variation of the 
hadronic input. For the middle values 
of all parameters specified above, 
we calculate $\overline{m}_s$ from (\ref{SR}) and 
obtain the central value presented in (\ref{interval}). 

The influence of the new $O(\alpha_s^4)$ correction 
on the sum rule is estimated
by repeating the calculation with the same 
input, but with the perturbative corrections
up to $O(\alpha_s^3)$. The result for the central value
turns out to be 
$\overline{m}_s(2 ~\mbox{GeV})=107$ MeV,  only 2 MeV larger   
than in (\ref{interval}). We also checked 
the quality of the OPE in the sum rule. Isolating 
the OPE part in (\ref{SR}),  that is, subtracting 
from l.h.s. of (\ref{SR}) the integral over $\rho^{(5)}_{OPE}(s) $ 
on r.h.s., we calculated the ratios 
$\tilde{r}_n^{(d=0,2)}(M^2,s_0)$ and $\tilde{r}^{(d=4,6)}(M^2,s_0)$ 
defined analogous to Eq.~(\ref{eq-rn}), where instead of 
$[\Pi^{(5)''}(M^2)]_{OPE}$  the contributions to the subtracted  
correlation function 
$$M^4[\Pi^{(5)''}(M^2)]_{OPE}-\int_{s_0}^\infty\!\! ds\,
e^{-s/M^2}\rho^{(5)}_{OPE}(s)$$ 
are substituted. We obtain $\tilde{r}_n^{(d=0,2)}(2.5~\mbox{GeV}^2,4.5~\mbox{GeV}^2)$=39.2\%, 26.1\%, 18.8\%, 10.6\%, 3.7\% for  $n=0,1,2,3,4$, 
respectively,  and  $\tilde{r}^{(d=4,6)}(2.5~\mbox{GeV}^2,4.5~\mbox{GeV}^2)=1.6\%$ revealing again a good convergence. 

Being numerically small, the $O(\alpha_s^4)$ correction
is nevertheless important for achieving a better stability 
with respect to the variation of the renormalization
scale $\mu$ entering the sum rule. To demonstrate that, we have calculated $\overline{m}_s(2 ~\mbox{GeV})$ from the sum rules with $O(\alpha_s^3)$ and $O(\alpha_s^4)$ accuracy, varying $\mu^2/M^2$ from 0.5 to 2.0.
%%%%%%%%%%%%%%%%%%%%%FIG.2 %%%%%%%%%%%%%%%%%%
\begin{figure}[t]
\begin{center}
{\includegraphics[scale=0.7]{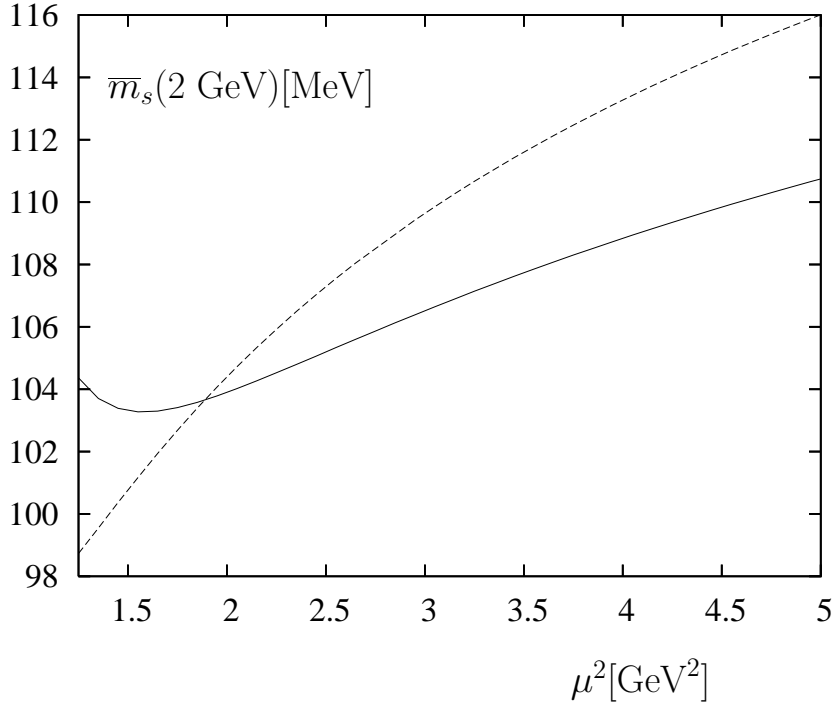}}
\end{center}
\caption{\it Strange quark mass at the scale 2 GeV, 
calculated from the sum rule (\ref{SR})
as a function of the renormalization scale $\mu^2$
in the correlation function, 
varying $\mu^2/M^2$ from 0.5 to 2.0 at $M^2=2.5 ~ GeV^2$.
The solid (dashed) line represents the result 
obtained with $O(\alpha_s^4)$ 
($O(\alpha_s^3)$) accuracy.}
\label{fig2}
\end{figure}
%%%%%%%%%%%%%%%%%%%%%%%%%
The results plotted in Fig.~2 clearly demonstrate the role
of the new $O(\alpha_s^4)$ correction in stabilizing 
the scale-dependence.

To investigate separate theoretical uncertainties of the
sum rule (\ref{SR}) in more detail, we group them into two  categories:

a) uncertainties 
related to the input parameters in the correlation function
and in the sum rule: renormalization scale, difference
between using $\alpha_s(m_z)$ and $\alpha_s(m_\tau)$, Borel parameter, 
$u$- and $d$-quark masses, condensate densities;

b) uncertainties caused by the hadronic input: 
the decay constants $f_{K_{1}}$ and $f_{K_{2}}$, the effective threshold $s_0$
and the effect of switching off the total widths of $K_{1,2}$.

Varying the input parameters in the QCD part of the sum rule 
within the limits specified above,
we find that the 
the largest uncertainty in the category (a) is caused by the 
scale variation (see Fig.~2), whereas the sensitivity to the Borel  
mass variation is less than  $\pm 0.1 $ MeV, and the 
dependence on  the values of condensate densities is negligible. 
Adding separate uncertainties grouped in this category 
in quadrature, we obtain the interval 
$(\pm 6 ~\mbox{MeV})|_{param}$ included in (\ref{interval}).

To investigate the hadronic uncertainties grouped above 
in the category (b), 
the decay constants $f_{K_1}$ , $f_{K_2}$ and the threshold 
$s_0$ are varied one by one yielding $\pm 5$ MeV,
$\pm 3$ MeV  and $\pm 3$ MeV, respectively.  
To estimate the effect of multiparticle states in the sum rule 
the $m_s$ calculation is repeated with the total widths
of $K_{1,2}$ in (\ref{BW}) set to zero. The result for $\overline{m}_s$ 
increases by approximately 3 MeV,  which we conservatively 
interprete as an additional uncertainty $\pm 3$  MeV. All these 
individual uncertainties are again added in quadrature to produce
$(\pm 7~\mbox{MeV})|_{hadr}$  in (\ref{interval}).

The nonperturbative effects beyond OPE 
cannot be estimated without the knowledge 
of the instanton effects, which are beyond 
our scope and are therefore absent in (\ref{interval}).
According to the estimate \cite{KM01} one has to add
$\pm 9 $ MeV to the total budget of uncertainties. 
%%%%%%%%%%%%%%%%%%%%%%%%%%%%%%%%
\begin{table}[t]
\begin{center}
\begin{tabular}{| c | r | c |}
\hline
Method & ${\overline m_s}(2$ GeV)  & Ref. \\ 
& [MeV] &\\ 
\hline\hline
Pseudoscalar Borel sum rule &$ 105\pm 6\pm 7$ & This work \\
&$ 100\pm 6$ & \cite{KM01}(no ILM) \\
\hline
Pseudoscalar FESR  & $100\pm 12$& \cite{KM01}\\
\hline
Scalar Borel sum rule & $99\pm 16$&\cite{JOP02}\\
\hline
Vector FESR  & $139\pm 31$&\cite{EJS03}\\
%&&\\
\hline \hline
 & $81\pm 22$ &\cite{GJPPS04}\\
Hadronic $\tau$ decays& $96^{+5+16}_{-3-18}$&\cite{BCKtau05}\\
 & $104\pm 28  $             & \cite{Narison05} \\
\hline
$\tau$ decays $\oplus$ sum rules & $99\pm 28$ & \cite{Narison05}\\ 
\hline\hline 
                     & $97\pm 22$ & \cite{lat_DellaMorte05}\\
Lattice QCD ($n_f=2$)& 100 -130  & \cite{lat_Gockeler05}\\
                     &$101\pm 8 ^{+25}_{-0}$ & \cite{lat_Becirevic05}\\
\hline
                     & $76\pm 3\pm 7$  & \cite{lat_Aubin04}
\\ 
Lattice QCD ($n_f=3$)& $86.7\pm 5.9$  & \cite{lat_Ishikawa05}
\\ 
& $87\pm 4\pm 4$  & \cite{lat_Mason05}
\\ 
\hline\hline
PDG04 average &80 -130  &\cite{PDG}\\ 
\hline
\end{tabular}
\caption{\it Our estimate of $\overline{m}_s(2 \mbox{GeV)}$ 
compared with some recent determinations obtained 
with different methods. The error/uncertainty
identification in the results taken from the 
literature can be found in the 
corresponding papers. }
\label{table:ms}
\end{center}
\end{table}
%%%%%%%%%%%%%%%%%%
In Table 1 we compare our prediction with other 
determinations of $m_s$. The 
values of the $s$ quark mass  obtained from the correlation functions 
($\tau$ decays, Borel sum rules and FESR ) are consistent with 
each other and with the lattice QCD results within 
still large uncertainties, the lattice results with $n_f=3$ being  
systematically lower. Our estimate (\ref{interval})
is also consistent with the $s$-quark mass bound in 
$O(\alpha_s^4)$ obtained in \cite{as4}.

\section{Conclusion}

We have included the new $O(\alpha_s^4)$ correction
in the correlation function of the pseudoscalar strangeness-changing
quark currents and calculated the $s$-quark mass from the 
resulting Borel sum rule. 
In future the same analysis should be repeated for the scalar 
Borel sum rule, for both pseudoscalar and scalar FESR and for 
the sum rules with nonstrange light-quark currents, yielding
the $u$ and $d$ quark masses.

Our main intention here was to investigate  
the role of the $O(\alpha_s^4)$ terms in the OPE and in the 
sum rule. We have found that the new correction is comfortably small, 
making OPE in this channel very reliable.
Simultaneously, the addition of the $O(\alpha_s^4)$ 
contributions noticeably decreases
the renormalization scale-dependence of the resulting sum rule.

The QCD sum rules obtained on the basis of OPE 
still have a considerable room of improvement.
While the nonperturbative effects beyond OPE 
can be kept under control by choosing 
the virtuality (Borel parameter) scale large enough, 
and using ILM-type estimates, there is still a lack
of experimental information concerning the masses, total 
and partial widths of the excited kaon resonances.
The resonance  $K_1$ can be observed in 
$\tau \to K\pi\pi \nu_\tau $ decays, 
and both $K_{1}$ and $K_{2}$ probably also in hadronic 
$B$ decays where the currently available 
statistics allows to isolate many light-quark resonances
in the final states. 
With this information one would be able to built a 
hadronic spectral function in the pseudoscalar
channel which is less dependent on duality ansatz, 
so that the accuracy of the hadronic part 
of the sum rule eventually becomes closer to the  
high  precision achieved in the  QCD part.

\section*{Acknowledgements}

The authors are grateful to Johann K\"uhn and Thomas Mannel for useful discussions.
This work was supported by the Deutsche Forschungsgemeinschaft in the
Sonderforschungsbereich/Transregio SFB/TR-9 ``Computational Particle
Physics'' and with the project DFG KH205/1-1.

\end{document}